# METAL ENRICHMENT DISCRIMINATORS OF COLD FRONTS

R. Dupke

Dept. Astronomy, University of Michigan, Ann Arbor 48109

**ABSRACT**

Cold fronts are sharp surface brightness discontinuities characterized by a jump in gas temperature accompanied by a decline in X-ray surface brightness such that the gas pressure remains continuous across the front and, thus, these structures differ from bow shocks. Models suggest that cold fronts can be generated by external mechanisms involving the accretion of a subsystem with a remnant "cold core". However, internal mechanisms can also create cold fronts, such as gravitational scattering of subclumps or cD oscillation around the bottom of the potential well. These competing models for their formation can be discriminated through the measurements of the SN Type contamination across the front, which in turn can be determined from metal abundance ratios as measured from an ensemble of elements. Here we present the preliminary results of such analysis using a sample of clusters observed with Chandra.

## 1. INTRODUCTION & METHODS

Cold fronts have been originally interpreted as subsonic motion of accreted substructures (Markevitch et al. 2002). Although this explanation holds relatively well for clusters that have clear signs of merging, such as A3667 (Mazzotta et al. 2002) there are other alternative explanations that involve internal mechanisms. For example, cD oscillation in the bottom of the potential well or induced oscillation of the potential well itself, e.g. A496 (Dupke & White 2003), A1795 (Fabian et al. 2001; cf. Markevitch et al. 2001). Both types of models are justified theoretically (e.g. Bialek, Evrard and Mohr 2002, Lufkin et al. 1995, Tittley & Henriksen 2005 ).

The global enrichment history of clusters and groups is different (Finoguenov & Ponman 1999) and the internal distribution of metal enrichment is not flat in clusters (e.g. Dupke & White 2000; Allen et al. 2001); Typically, clusters have a higher SN Ia Fe mass contamination near the core than in the outer parts. This fact provides a way to test for competing cold front scenarios by looking at gradients in abundance ratios measured across the front. If the mechanism is external, such as a sub-clump accretion, one should find an enrichment discontinuity across the front. If no discontinuity is found models based on internal mechanisms are favored. Dupke & White (2003) used this method with Abell 496 and found no evidence of a chemical discontinuity across the front within the errors.

We here extend their analysis to a larger sample of "cold front" clusters to determine the frequency and overall properties of different mechanisms that generate cold fronts. The data presented here were reduced with **Ciao** 3.2.0 with **CALDB** 3.0 using the standard procedure and blank-sky background. Here we show the results of spectral fittings with **XSPEC** V11.3.1 using an absorbed **VAPEC** thermal emission model. The maps shown below are colored contour maps. The color steps were chosen in a way as to approximate the 1σ errors. Below, for illustration, we show the projected distribution of entropy $T/n^{2/3}$ (K cm$^2$)), which is a good tracer of cold fronts (marked with white arrows) and Si/Fe for A496 and A4059 and compare them to the merging cluster A576, for which there are strong indications of a strong line of sight merger with a NW-SE inclination of ~9º (Dupke, Mirabal and Bregman 2006).

## 2. RESULTS

In the "remnant scenario" model for cold fronts, the expected distribution of abundance ratios should be similar to what is observed for the merging cluster A576 (last row), where a clear Si/Fe discontinuity is seen at the cluster core. The results of a deeper observation of A496 do not indicate any obvious chemical gradient using 4 well define abundance ratios including Si, S, O and Fe in any of its 3 confirmed cold fronts. Furthermore, similar analysis performed in 3 other clusters with cold fronts (2A0335, A2052, A4059) show similar results. The results are in agreement with models that use internal mechanisms to form cold fronts. Despite the presence of chemical gradients near the core of some clusters (e.g. A4059 below), the distribution of abundance ratios do not show any systematic connection to the surface discontinuities as it should be expected and is yet to be detected, in the "remnant core" scenario. Therefore, association of cold fronts to ICM bulk motions may not be justified for a significant fraction of clusters.

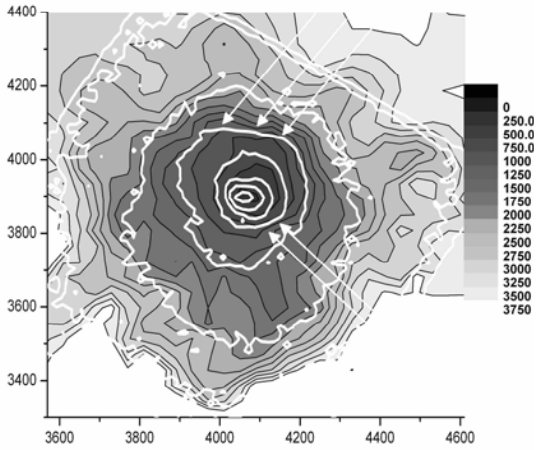
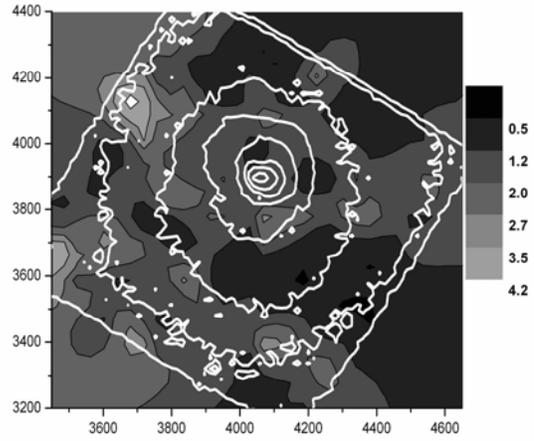
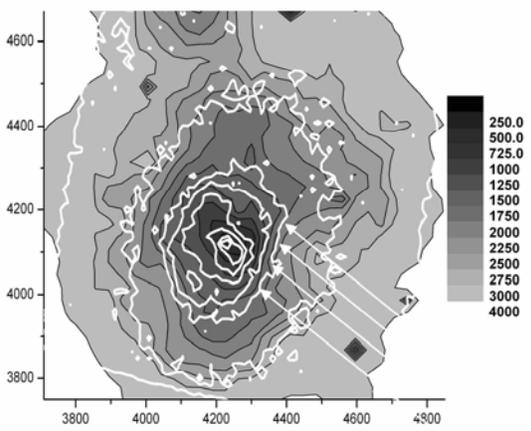
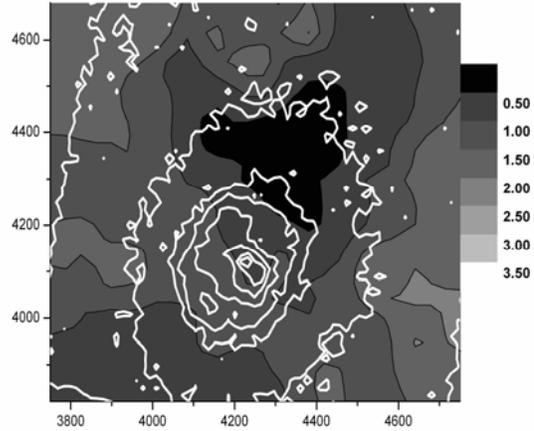
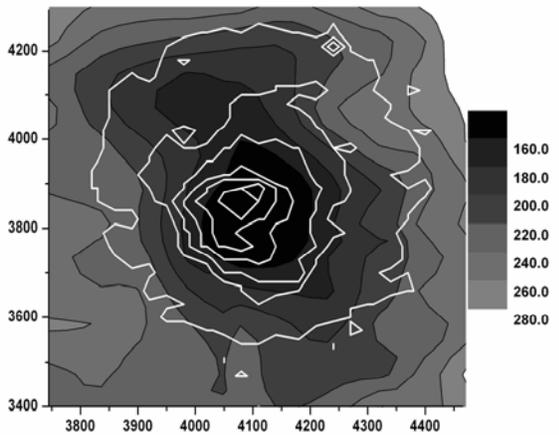
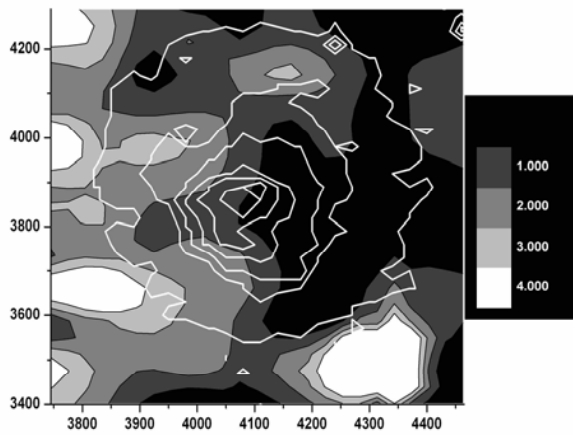